\documentstyle[prd,aps]{revtex}
\begin{document}


\title{
{\bf NOTE ON SIGNATURE CHANGE AND COLOMBEAU THEORY}
}

\author{Tevian Dray}
\address{
Department of Mathematics, Oregon State University,
  Corvallis, OR  97331, USA \\
{\tt tevian{\rm @}math.orst.edu}
}

\author{George Ellis}
\address{
Department of Mathematics and Applied Mathematics, University of Cape 
Town, Rondebosch 7701, SOUTH AFRICA \\
{\tt ellis{\rm @}maths.uct.ac.za} \\[3mm]
}

\author{Charles Hellaby}
\address{
Department of Mathematics and Applied Mathematics, University of Cape 
Town, Rondebosch 7701, SOUTH AFRICA \\
{\tt cwh{\rm @}maths.uct.ac.za} \\[3mm]
}

\date{July 21, 2000}

\maketitle

\begin{abstract}
Recent work alludes to various `controversies' associated with
signature change in general relativity.  As we have argued
previously, these are in fact disagreements about the (often
unstated) assumptions underlying various possible approaches.
The choice between approaches remains open.
\end{abstract}

\section{Introduction}

In a recent paper \cite{Mansouri}, Mansouri and Nozari (MN) discuss
signature change in General Relativity Theory from the point of view
of Colombeau's generalised functions.  They then claim that this
``solves the controversy in the literature'' regarding the
``vanishing of ... Einstein's equation on a surface of signature
change.''  We argue that there is in fact no mathematically based
issue here that can be `resolved' in this way.  Rather, there is no
clear choice of what should qualify as `Einstein's equations' in this
generalised context, and several options are open, even when Colombeau 
theory is used.  One can make various mathematical proposals in this 
regard; which of these, if any, have anything to do with real physics is 
unclear, and is open to debate.  We also point out that a reasonable 
interpretation of Mansouri and Nozari's equations leads to boundary 
conditions at a signature change which support the choice made by Dray et 
al.~\cite{Particle,Signature,Failure} and Ellis and 
coworkers~\cite{Ellis1,Ellis2,Carfora} in the original signature change 
`controversy'~\cite{HaywardI,Boundary,HaywardComment,Reply}.  To 
illustrate the point, we append an example with 
 non-zero extrinsic curvature, contrasting with MN's example that has zero 
extrinsic curvature at the signature change. 

\section{Einstein's Equations and Boundary Conditions}

It is not possible for a nondegenerate metric to change signature,
so signature change inevitably involves degenerate metrics.  But
Einstein's equations are usually derived under the assumption that
the metric is nondegenerate, so attempts to apply the Einstein
equations through such a signature change run into problems.  Several
attempts have been made to address this apparent conflict by making
generalisations of the Einstein equations, with varying results.  An
alternative is simply to apply the standard Einstein equations on
each side of the surface where the signature changes, and to impose
suitable boundary conditions at that surface.

Thus, in the absence of a clear statement of just what Einstein's
equations are in these circumstances, authors have argued in favor of
particular boundary conditions, which should hold on a surface of
signature change.  The two basic choices are for the extrinsic
curvature to be continuous but not necessarily
zero~\cite{Particle,Signature,Failure,Ellis1,Ellis2}, or for it to be
continuous and to vanish there \cite{Hayward}.  One can also consider
discontinuities in the extrinsic curvature \cite{Carfora}, but
attempts to relate these to a distributional matter source at the
boundary require some form of field equations valid on the surface.

Several authors have given derivations of generalised forms of the
Einstein's equations in the presence of signature change.  Kossowski
and Kriele \cite{KK} assume that the standard form of Einstein's
equations written in a specific way should continue to hold in the
presence of signature change, and show that the absence of a surface layer 
(i.e. a distributional term in the matter at the boundary) then forces the 
extrinsic curvature to vanish there.  Dray \cite{Einstein,Gravity} derives 
(a different version of) Einstein's equations from a variational 
principle, and shows that the absence of a surface layer in this case 
forces the extrinsic curvature to be continuous but not necessarily zero
 --- the Darmois boundary conditions \cite{Darmois}.  Thus, each of the 
basic boundary conditions follows from forms of the Einstein field 
equations, which in turn can be derived from suitable starting points; 
different starting points give different results. 

Mansouri and Nozari \cite{Mansouri} repeat the standard calculation
of the Einstein tensor using Colombeau's generalised functions, and
claim (equations (33)--(36)) that it is proportional to the
discontinuity in the extrinsic curvature.  Thus, also in this approach, 
the absence of a surface layer at the boundary forces the extrinsic 
curvature to be continuous, but not necessarily zero.  This is made clear 
by the example in the appendix.

Furthermore, the proportionality factor in the Mansouri and Nozari
result for the energy momentum tensor on the change surface (equation
(57)) depends on the regularisation used in the Colombeau version of
the Dirac delta function at the signature change surface.  This is
rather awkward, and can only be avoided if the extrinsic curvature is
continuous, which implies that this energy-momentum tensor is zero.
It thus appears that a reasonable interpretation of the Mansouri and
Nozari approach is that it is only possible in a parameterisation
invariant way if there is no distributional term in the matter at the 
boundary, unlike the other derivations just cited, and in apparent 
contradiction to claims in their paper (page 266). 

Finally, although these calculations if anything support our previous
position rather than others that have been put forward (indeed the paper 
appears to be incorrect when it states on page 255 that it supports the 
result of their reference 18 --- our \cite{HaywardComment}), we record an 
uneasiness with its methods due to some problems with the `microstructure' 
embodied in the Colombeau approach, adopted in order to avoid the usual 
problems associated with products of distributions. Kamleh~\cite{Kamleh} 
has recently given a rigorous treatment of signature change using 
Colombeau's generalised functions, and concludes that ``the Colombeau 
algebra is ... unable to settle the dispute over the nature of the 
junction conditions.'' 

\section{Conclusion}

As we have argued elsewhere \cite{KKcomment}, one must ultimately
choose what one means by Einstein's equations in the presence of
signature change; this does not follow from the standard equations,
which are only valid for a regular metric, and so involves some
choice as to how to generalise the usual equations in a way that
covers this situation.  Different mathematical choices lead to
different boundary conditions, and may be relevant to different
physical situations.  The paper by Mansouri and Nozari
\cite{Mansouri} gives a derivation of one possible version of
Einstein's equations within the framework of Colombeau's generalised
functions.  Other versions, together 
with suitable derivations, exist, as shown by Kamleh~\cite{Kamleh}.  
Thus, attempts to label any specific choice as `the' Einstein
equations are necessarily based in rhetoric rather than physics or
mathematics.  Ultimately, one must choose which version one prefers;
such a choice is not forced on one by the mathematics of the
situation, and what experimental physics  may imply is as yet unknown.

The example provided shows that MNs results do not support the view of 
\cite{HaywardI,HaywardComment,Hayward,KK}.


\appendix

\section{Contrasting Example}

 We start from the de Sitter metric, eq (59) of MN, and use their methods 
and notation, 
 \begin{equation}
   ds^2 = -f(t) \, dt^2 + a^2(t) \left( d\chi^2 + \sin^2 \chi ( d\theta^2 
+ \sin^2 \theta \, d\phi^2) \right) ~~.
 \end{equation}
 In a Euclidean region, where $f < 0$, 
 \begin{equation}
   a_E = \alpha_- \, \cos \left( \frac{t}{\alpha_-} \right)
   ~~,~~~~~~~~~~~~
   \dot{a}_E =- \sin \left( \frac{t}{\alpha_-} \right) ~~,
 \end{equation}
 and in a Lorentzian region, $f > 0$, 
 \begin{equation}
   a_L = \alpha_+ \, \cosh \left( \frac{t}{\alpha_+} \right)
   ~~,~~~~~~~~~~~~
   \dot{a}_L =\sinh \left( \frac{t}{\alpha_+} \right) ~~.
 \end{equation}
 Now, according to MN's equations (53)-(56), there would be no surface 
layer at a signature change if $[\dot{a}] = 0$.  As MN say on p268, it 
remains to be checked whether signature changes at times other than the 
maximum (minimum) in $a(t)$ on the Euclidean (Lorentzian) side can be free 
of surface layers.  We do that here. 

 We insert a signature change at some time $t_-$ on the Euclidean side and 
time $t_+$ on the Lorentzian side, and then shift the origin of time on 
both sides to make $t = 0$ there.  Specifically, 
 \begin{equation}
   f = - \theta(-t) + \theta(t)
   ~~,~~~~~~~~~~~~
   a^2 = a_E^2(t_- + t) \, \theta(-t) + a_L^2(t_+ + t) \, \theta(t) ~~,
 \end{equation}
 and we have no surface layer at $t = 0$ if
 \begin{equation}
   a_- = a_E(t_-) =a_- = a_L(t_+)
   ~~,~~~~~~~~~~~~
   \dot{a}_- = \dot{a}_E(t_-) =\dot{a}_+ = \dot{a}_L(t_+) ~~.
 \end{equation}
 This is achieved by solving
 \begin{equation}
   \alpha_- \, \cos \left( \frac{t_-}{\alpha_-} \right) = 
      \alpha_+ \, \cosh \left( \frac{t_+}{\alpha_+} \right)
   ~~~~~~\mbox{and}~~~~~~
   - \sin \left( \frac{t_-}{\alpha_-} \right) = 
      \sinh \left( \frac{t_+}{\alpha_+} \right) ~~,
 \end{equation}
 for which a couple of specific numerical solutions are:
 \begin{eqnarray}
   \mbox{choosing~~~~} && \alpha_- = 3 ~~,~~~~~~
      t_- = -0.5 ~~~~\rightarrow~~~~
      \alpha_+ = 2.918540931 ~~,~~~~~~
      t_+ = 0.4819808451 \\
   && ~~~~\rightarrow~~~~
      a_\pm = 2.958429695 ~~,~~~~~~
      \dot{a}_\pm = 0.1658961327 ~~; \\
   \mbox{choosing~~~~} && \alpha_- = 7 ~~,~~~~~~
      t_- = -1 ~~~~\rightarrow~~~~
      \alpha_+ = 6.859521334 ~~,~~~~~~
      t_+ = 0.9733324138 \\
   && ~~~~\rightarrow~~~~
      a_\pm = 6.928692823 ~~,~~~~~~
      \dot{a}_\pm = .1423717298 ~~.
 \end{eqnarray}
 MN's example had $t_- = 0$, $t_+ = 0$, $\alpha_+ = \alpha_-$.  Clearly, 
signature changes without surface layers are possible at any time.  This 
applies not only to the de Sitter case, but almost any RW model, and 
indeed more generally%
 \footnote{
 We note that the cosmological constant jumps from $3/\alpha_-^2$ to 
$3/\alpha_+^2$ in the above example, but this is not a problem, as 
discontinuities in the matter occur with the usual (Lorentzian to 
Lorentzian) boundary conditions, whereas at a signature change the whole 
nature of physics changes and causality suddenly appears.  Indeed a jump 
is a much weaker singularity than a surface layer, which MN allow (top 
half of p266). 
 }%
 .

 For comparison, the Darmois-Israel conditions require continuity of the 
intrinsic metric and extrinsic curvature on the change surface, 
 \begin{equation}
   h_{ij} = a^2 \, {\rm diag}(1, \sin^2 \chi, \sin^2 \chi \, \sin^2 
\theta) ~~,~~~~~~~~~~~~
   K_{ij} = \frac{a \dot{a}}{\sqrt{|f|}\;} \, {\rm diag}(1, \sin^2 
\chi, \sin^2 \chi \, \sin^2 \theta) ~~.
 \end{equation}
 i.e. $[a] = 0$ and $[\dot{a}/\sqrt{|f|}\;] = 0$.  Provided a continuous 
$a$ is implicit and $|f_+| = |f_-| > 0$, there is no surface layer, in 
agreement with the above%
 \footnote{
 It should be emphasised that the surface effects found in \cite{Failure} 
are delta functions in the conservation laws, not in the Einstein/matter 
tensor.  
 }%
 .


\begin{references}

\bibitem{Mansouri}
Reza Mansouri and Koruosh Nozari,
{\it A New Distributional Approach to Signature Change},
Gen.\ Rel.\ Grav.\ {\bf 32}, 253--269 (2000).

\bibitem{Particle}
Tevian Dray, Corinne A. Manogue, and Robin W. Tucker,
{\it Particle Production from Signature Change},
Gen.\ Rel.\ Grav.\ {\bf 23}, 967 (1991).

\bibitem{Signature}
Tevian Dray, Corinne A. Manogue, and Robin W. Tucker,
{\it The Scalar Field Equation in the Presence of Signature Change},
Phys.\ Rev.\ {\bf D48}, 2587 (1993).

\bibitem{Failure}
Charles Hellaby and Tevian Dray,
{\it
Failure of standard conservation laws at a classical change of 
signature},
Phys.\ Rev.\ {\bf D49}, 5096--5104 (1994).

\bibitem{Ellis1}
G Ellis, A Sumeruk, D Coule, C Hellaby,
{\it Change of Signature in Classical Relativity},
Class.\ Quant.\ Grav.\ {\bf 9}, 1535 (1992).

\bibitem{Ellis2}
G F R Ellis,
{\it Covariant Change of Signature in Classical Relativity},
Gen.\ Rel.\ Grav.\ {\bf 24}, 1047 (1992).

\bibitem{Carfora}
Mauro Carfora and George Ellis,
{\it The Geometry of Classical Change of Signature},
Intl.\ J. Mod.\ Phys.\ {\bf D4}, 175 (1995).

\bibitem{HaywardI}
Sean A. Hayward, {\it Weak Solutions Across a Change of Signature},
Class.\ Quantum Grav.\ {\bf 11}, L87 (1994).

\bibitem{Boundary}
Tevian Dray, Corinne A. Manogue, and Robin W. Tucker,
{\it Boundary Conditions for the Scalar Field
 in the Presence of Signature Change},
Class.\ Quantum Grav.\ {\bf 12}, 2767-2777 (1995).

\bibitem{HaywardComment}
Sean A. Hayward,
{\it Comment on
``Failure of Standard Conservation Laws at a Classical Change of 
Signature''},
Phys.\ Rev.\ {\bf D52}, 7331--7332 (1995).

\bibitem{Reply}
Charles Hellaby and Tevian Dray,
{\it Reply Comment: Comparison of Approaches to Classical Signature 
Change},
Phys.\ Rev.\ {\bf D52}, 7333--7339 (1995).

\bibitem{Hayward}
Sean A. Hayward,
{\it Signature Change in General Relativity},
Class.\ Quant.\ Grav.\ {\bf 9}, 1851 (1992);
erratum: Class.\ Quant.\ Grav.\ {\bf 9}, 2543 (1992).

\bibitem{KK}
M. Kossowski and M. Kriele,
{\it Smooth and Discontinuous Signature Type Change in General 
Relativity},
Class.\ Quant.\ Grav.\ {\bf 10}, 2363 (1993).

\bibitem{Einstein}
Tevian Dray,
{\it Einstein's Equations in the Presence of Signature Change},
J.~Math.\ Phys.\ {\bf 37}, 5627--5636 (1996).

\bibitem{Gravity}
Tevian Dray, George Ellis, Charles Hellaby, and Corinne A. Manogue,
{\it Gravity and Signature Change},
Gen.\ Rel.\ Grav.\ {\bf 29}, 591--597 (1997).

\bibitem{Darmois}
G. Darmois,
{\bf M\'emorial des Sciences Math\'ematiques},
Fascicule 25, Gauthier-Villars, Paris, 1927.

\bibitem{Kamleh}
Waseem Kamleh,
{\it Signature Changing Space-times and the New Generalised 
Functions},
{\tt gr-qc/0004057}.

\bibitem{KKcomment}
Tevian Dray and Charles Hellaby,
{\it Comment on
`Smooth and Discontinuous Signature Type Change in General 
Relativity'},
Gen.\ Rel.\ Grav.\ {\bf 28}, 1401--1408 (1996).

\end{references}
\end{document}